\documentclass[11pt]{article}

\usepackage{epsfig}

\newcommand{\gsim}{\mbox{\raisebox{-1.ex}{$\stackrel
     {\textstyle>}{\textstyle\sim}$}}}
\newcommand{\lsim}{\mbox{\raisebox{-1.ex}{$\stackrel
     {\textstyle<}{\textstyle \sim}$}}}
\newcommand{\square}{\kern1pt\vbox{\hrule height
1.2pt\hbox{\vrule width 1.2pt\hskip 3pt
   \vbox{\vskip 6pt}\hskip 3pt\vrule width 0.6pt}\hrule
height 0.6pt}\kern1pt}

\newcommand{\beq}{\begin{equation}}
\newcommand{\eeq}{\end{equation}}
\newcommand{\beqa}{\begin{eqnarray}}
\newcommand{\eeqa}{\end{eqnarray}}
\newcommand{\bea}{\begin{array}}
\newcommand{\ena}{\end{array}}

\def\bea{\begin{eqnarray}}
\def\eea{\end{eqnarray}}
\def\beq{\begin{equation}}
\def\eeq{\end{equation}}

\def\be{\begin{equation}}
\def\ee{\end{equation}}

\def\5{\overline 5}
\begin{document}

\thispagestyle{empty}

\begin{center}
\begin{center}

\vspace{1.7cm}

{\LARGE\bf Introductory Review of Cosmic Inflation}

\end{center}

\vspace{1.4cm}

{\large Shinji 
Tsujikawa\footnote{email: 
shinji@resceu.s.u-tokyo.ac.jp,
shinji@gravity.phys.waseda.ac.jp}}\\

\vspace{1.2cm}

{\em Research Center for the Early Universe, University of Tokyo,} \\
{\em Hongo, Bunkyo-ku, Tokyo 113-0033, Japan}

\vspace{.3cm}

\end{center}

\vspace{0.8cm}

\vspace{3cm}

\centerline{\Large Abstract}
\vspace{3 mm}
\begin{quote}\small
These lecture notes provide an introduction to cosmic inflation.  In 
particular I will review the basic concepts of inflation, generation of 
density perturbations, and reheating after inflation.
\end{quote}

\vfill 
\newpage

\section{Preface}

These lecture notes are based on invited lectures at The Second Tah Poe 
School on Cosmology ``Modern Cosmology",
Naresuan University, Phitsanulok, Thailand, April 17 -25, 2003.
See the web page 

{\tt http://www.tech.port.ac.uk/staffweb/gumjudpb/TPC2FIRSTpage.html} \\
for the details of this school and conference.

\section{Introduction}

The proposal of General 
Relativity by Einstein in 1915 made it possible to discuss the structure of 
spacetime and the evolution of the universe in terms of physical laws.  In 
1922 Friedmann found the existence of expanding/collapsing cosmological  
solutions by solving the Einstein field equations.  In 1929 Hubble 
discovered the expansion of the universe by the observations in the 
redshift of galaxies, as the Einstein theory predicts.  In 1946 Gamov and 
his collaborators showed that the universe must begin in a very hot and 
dense state from the theory of nucleosynthesis.  They also predicted that 
the present universe should be filled with the microwaves with black body 
radiations.  In 1965 Penzias and Wilson discovered microwave background 
radiations which coincide well with theoretical predictions by Gamov {\it et 
al.}.  These strong observational evidences have made people believe that the 
universe started out from a hot and dense state, which is called as the 
standard big-bang model.

In standard big-bang cosmology the state of the universe is characterized
by the radiation-dominated or the matter-dominated stage. 
These correspond to the decelerated expansion of the universe
where the second derivative of the scale factor is negative.
Meanwhile this decelerated 
expansion is not sufficient to solve a number of cosmological problems such 
as flatness and horizon problems which plagues in the standard big-bang 
scenario\footnote{In the next section I will explain how inflation solves 
these problems.}.  In order to overcome these fundamental problems, it is 
required to consider an epoch of accelerated expansion in the early 
universe, {\it i.e.}, {\it inflation}.
  
The basic ideas of inflation were originally proposed 
by Guth \cite{Guth} and Sato \cite{Sato} independently in 1981, 
which is now termed as {\it old inflation}\footnote{Note that a specific
version of the inflationary scenario--so called $R^2$ inflation-- was 
proposed by Starobinsky one year earlier \cite{Star}.  Nevertheless this did 
not explicitly point out the virtue of inflation as in the paper of Guth.}.  
This corresponds to the de-Sitter inflation which makes use of the 
first-order transition to true vacuum.  However, it has a serious 
shortcoming that the universe becomes inhomogeneous by the bubble 
collision soon after the inflation ends.  The revised version was proposed 
by Linde \cite{Linde82}, and Albrecht and Steinhardt \cite{AS82} in 1982, 
which is dubbed as {\it new inflation}.  This corresponds to the slow-roll 
inflation with the second-order transition to true vacuum.  Unfortunately 
this scenario also suffers from a fine-tuning problem of spending enough 
time in false vacuum to lead to sufficient amount of inflation.  In 1983 
Linde \cite{Linde83} considered the variant version of the slow-roll 
inflation called {\it chaotic inflation}, in which initial conditions of 
scalar fields are chaotic.  According to this model, our homogeneous and 
isotropic universe may be produced in the regions where inflation occurs 
sufficiently.  While old and new inflation models are based on the 
assumption that the universe was in a state of thermal equilibrium from the 
beginning, chaotic inflation can occur even without such an assumption.  In 
addition chaotic inflation can start out in the regime close to Planck 
density, thereby solving the problem of initial conditions.

Many kinds of inflationary models have been constructed in these twenty 
years (see {\it e.g.,} ref.~\cite{Linde,LL}).  In particular, the recent trend 
is to construct consistent models of inflation based on superstring or 
supergravity models.  See ref.~\cite{LR99} for a review in such a direction.
  
The inflationary paradigm not only provides an excellent way in solving 
flatness and horizon problems but also generates density perturbations as 
seeds for large scale structure in the universe \cite{inper}.  In fact 
inflation provides a causal mechanism to generate nearly scale-invariant 
spectra of cosmological perturbations.  Quantum fluctuations of the field 
driving inflation--called {\em inflaton}-- are 
typically frozen by the accelerating expansion when the scales of 
fluctuations leave the Hubble radius.  Long after the inflation ends, the 
scales cross inside the Hubble radius again.  Thus perturbations imprinted 
during inflation can be the origin of large-scale structure in the 
universe.  In fact temperature anisotropies observed by the COBE satellite 
in 1992 exhibit nearly scale-invariant spectra as predicted by the 
inflationary paradigm.  Recent observations of WMAP also show strong 
evidence for inflation \cite{Bennett:2003bz}.

\section{The standard big-bang cosmology and its problems}

\subsection{The standard big-bang cosmology}

The standard big-bang cosmology is based on 
the cosmological principle \cite{LL}, which means that the universe is
homogeneous and isotropic on large distance. 
Then the metric takes the Friedmann-Robertson-Walker (FRW) form: 
\beqa 
ds^2=g_{\mu\nu}dx^{\mu}dx^{\nu}= -dt^2+a^2(t)\left[\frac{dr^2}{1-Kr^2}+ 
r^2(d\theta^2+\sin^2\theta d \phi^2) \right].
\label{2_1_1}
\eeqa 
Here $a(t)$ is the scale factor with $t$ being the cosmic time. 
The constant $K$ is the spatial curvature, where positive, zero, and negative 
values correspond to closed, flat, and open universes, respectively.

The evolution of the universe is dependent on the material within it.  This is 
characterized by the equation of state between the energy density $\rho(t)$ 
and the pressure $p(t)$.  
Typical examples are : 
\beqa 
p&=&\rho/3\,,~~~~~~{\rm radiation}\,, \\
p&=&0\,,~~~~~~~~~{\rm dust}\,.
\label{2_1_2}
\eeqa 
In order to know the dynamical evolution of the universe, it is required to 
solve the Einstein equations in General Relativity.  
The Einstein equations are expressed as \cite{Weinberg,Gravitation} 
\beqa 
G_{\mu \nu} \equiv R_{\mu\nu}-\frac12 g_{\mu\nu} R= 
8\pi G T_{\mu\nu}-\Lambda 
g_{\mu\nu},
\label{2_1_3}
\eeqa 
where $R_{\mu\nu}$, $R$, $T_{\mu\nu}$, and 
$G$ are the Ricci tensor, Ricci scalar, energy momentum tensor, 
gravitational constant, respectively.  The Planck energy, 
$m_{\rm pl}=1.2211 \times 10^{19}$\,GeV, is related with $G$ through the 
relation $m_{\rm pl}=(\hbar c^5/G)^{1/2}$.  Here
$\hbar$ and $c$ are the Planck's constant and the speed of light, 
respectively.
Hereafter we use the units $\hbar=c=1$.  
$\Lambda$ is a cosmological constant 
originally introduced by Einstein.

For the background metric (\ref{2_1_1}) with a negligible cosmological 
constant, the Einstein equations (\ref{2_1_3}) yield 
\beqa 
H^2=\frac{8\pi}{3m_{\rm pl}^2}\rho-\frac{K}{a^2}\,,
\label{2_1_4}
\eeqa 
\beqa
\dot{\rho}+3H(\rho+p)=0\,,
\label{2_1_5}
\eeqa 
where a dot denotes the derivative with respect to $t$, and $H \equiv 
\dot{a}/a$ is the Hubble expansion rate.  Eqs.~(\ref{2_1_4}) and 
(\ref{2_1_5}) are so called the Friedmann and Fluid equations, 
respectively.  Combining these relations gives the following equation, 
\beqa 
\frac{\ddot{a}}{a}=-\frac{4\pi}{3m_{\rm pl}^2} (\rho+3p)\,.
\label{2_1_6}
\eeqa 
The Friedmann equation (\ref{2_1_4}) can be rewritten as
\beqa
\Omega-1=\frac{K}{a^2H^2}\,,
\label{2_1_7}
\eeqa 
where 
 \beqa
\Omega \equiv \frac{\rho}{\rho_c},~~~~~{\rm with}
~~~~~\rho_c \equiv \frac{3H^2m_{\rm pl}^2}{8\pi}\,.
\label{2_1_8}
\eeqa 
Here the density parameter $\Omega$ is the ratio of the energy density to the 
critical density. 
When the spatial geometry is flat ($K=0$), 
the solutions for Eqs.~(\ref{2_1_4}) and (\ref{2_1_5}) are 
\beqa 
{\rm Radiation~~dominant}:~~~~~a \propto t^{1/2}\,,~~~~~
\rho \propto a^{-4}\,, \\
{\rm Dust~~dominant}:~~~~~a \propto t^{2/3}\,,~~~~~\rho \propto a^{-3}\,.
\label{2_1_9}
\eeqa 
In these simple cases, the universe is 
expanding deceleratedly ($\ddot{a}<0$) as confirmed by Eq.~(\ref{2_1_6}).

\subsection{Problems of the standard big-bang cosmology}

\subsubsection{Flatness problem}

In the standard big-bang theory with $\ddot{a}<0$, 
the $a^2H^2 (=\dot{a}^2)$ term in 
Eq.~(\ref{2_1_7}) always decreases.  This indicates that $\Omega$ 
tends to shift away from unity with the expansion of the universe.  
However, since present observations suggest that $\Omega$ is within the 
order of magnitude of one \cite{Bennett:2003bz}, $\Omega$ needs to be very 
close to one in the past.  For example, we require $|\Omega-1| < {\cal 
O}(10^{-16})$ at the epoch of nucleosynthesis \cite{LL,Liddle} and 
$|\Omega-1| < {\cal O}(10^{-64})$ at the Planck epoch \cite{Riotto}.  This 
is an extreme fine-tuning of initial conditions.  Unless initial conditions 
are chosen very accurately, the universe soon collapses, or expands quickly 
before the structure can be formed.  This is so called the flatness 
problem.

\subsubsection{Horizon problem}

Consider a comoving wavelength, $\lambda$, and also a physical wavelength, 
$a\lambda$, which is inside the Hubble radius, $H^{-1}$ 
({\it i.e.,}~$a\lambda~\lsim~H^{-1}$).  The standard big-bang cosmology is 
characterized by the cosmic evolution of $a \propto t^p$ with $0<p<1$.  In 
this case the physical wavelength grows as $a\lambda \propto t^p$, whereas 
the Hubble radius evolves as $H^{-1} \propto t$.  Therefore the physical 
wavelength becomes much smaller than the Hubble radius with the passage of 
time.  This means that the region where the causality works eventually 
becomes the only small fraction of the Hubble radius.

To be more precise, let us first define the particle horizon $D_H(t)$ where 
the light travels from the beginning of the universe, $t=t_*$, \beqa 
D_H(t)=a(t)d_H(t)\,,~~~~{\rm with }~~~~ d_H(t)=\int_{t_*}^t 
\frac{dt}{a(t)}\,.
\label{2_1_11}
\eeqa 
Here $d_H(t)$ corresponds to the comoving distance.  Setting $t_*=0$, we find 
$d_H(t)=3t$ in the matter-dominant era.  We observe photons in the 
Cosmic Microwave Background (CMB) which are emitted at the time of {\it 
decoupling}. 
The particle horizon at decoupling, 
$D_H(t_{\rm dec})=a(t_{\rm dec})d_H(t_{\rm dec})$, 
corresponds to the region where photons could have contacted causally at 
that time.  The ratio of $d_H(t_{\rm dec})$ to the particle horizon today, 
$d_H(t_0)$, where $t_0$ is the present time, is estimated as \beqa 
\frac{d_H(t_{\rm dec})}{d_H(t_0)} \approx \left(\frac{t_0}{t_{\rm 
dec}}\right)^{1/3} \approx \left(\frac{10^5}{10^{10}}\right)^{1/3} \approx 
10^{-2}.
\label{2_1_12}
\eeqa 
This result implies that the causality regions of photons
are restricted to be small. In fact the surface of the last scattering 
surface only corresponds to the angle of order $1^{\circ}$. 
Observationally, however, we see photons which thermalize to the same 
temperature in all regions in the CMB sky.  This is so called the horizon 
problem.

\subsubsection{The origin of large-scale structure in the universe}

The COBE satellite observes the anisotropies in 
the last scattering surface, 
whose amplitudes are small and close to scale-invariant.  These fluctuations 
spread so large a scale that it is practically impossible to generate them 
between the big bang and the time of the last scattering in the standard 
cosmology.  This problem is almost equivalent to the horizon problem mentioned 
above, {\it i.e.}, the standard cosmology can not provide satisfactory 
explanation for the origin of large-scale structure.

\subsubsection{Monopole problem}

According to the view point of particle physics, 
the breaking of supersymmetry leads to the production of many unwanted relics 
such as monopoles, cosmic strings, and topological defects \cite{Linde}.  
The string theories also predict supersymmetric particles such as 
gravitinos, Kaluza-Klein particles, and moduli fields.

If these particles exist in the early stage of the universe,
the energy densities of them decrease as a matter component ($\sim a^{-3}$).
Since the radiation energy density decreases as $\sim a^{-4}$ in the 
radiation-dominant era, these massive relics could be the dominant materials 
in the universe, which contradicts with observations.  This problem is 
generally called as the monopole problem.

\section{Idea of inflationary cosmology} 

The problems in the big-bang cosmology lie in the fact that the universe 
always exhibits the decelerating expansion.  Let us assume the existence of 
a stage in the early universe with an accelerated expansion of 
the universe, {\it i.e.}, 
\beqa 
\ddot{a}>0\,.
\label{2_2_1}
\eeqa
{}From the relation (\ref{2_1_6}) this corresponds to the condition 
\beqa 
\rho+3p<0\,.
\label{2_2_2}
\eeqa
The condition (\ref{2_2_1}) essentially means that 
$\dot{a}\,(=aH)$ increases during inflation. Then the
comoving Hubble radius, $(aH)^{-1}$, decreases in the inflationary phase.  
This property is the key point to solve the cosmological puzzles in the 
standard big-bang cosmology as I will show below.

\subsection{Flatness problem}

Since the $a^2H^2$ term in Eq.~(\ref{2_1_7}) increases during 
inflation, $\Omega$ rapidly approaches unity.  After the inflationary period 
ends, the evolution of the universe is followed by the conventional big-bang 
phase and $|\Omega-1|$ begins to increase.  In spite of this, as long as 
the inflationary expansion occurs sufficiently and makes $\Omega$ very 
close to one, $\Omega$ stays of order unity even in the present epoch.

\subsection{Horizon problem}

Since the scale factor evolves as $a \propto t^{p}$ with $p>1$
during inflation, the physical wavelength, $a\lambda$, grows faster 
than the Hubble radius, $H^{-1} (\propto t)$.
Therefore the physical wavelength is pushed outside the Hubble radius
during inflation.  This means that the region where the causality works 
is stretched on scales much larger than the Hubble radius,
thus solving the horizon problem.

Of course the Hubble radius begins to grow faster than the physical 
wavelength after inflation (radiation/matter dominant era).
In order to solve the horizon problem, it is required that the following 
condition is satisfied for the comoving particle horizon: 
\beqa 
\int_{t_*}^{t_{\rm dec}}\frac{dt}{a(t)} \gg \int_{t_{\rm 
dec}}^{t_0}\frac{dt}{a(t)}\,.
\label{2_2_3}
\eeqa
This implies that the comoving distance that photons can travel before 
decoupling needs to be much larger than that after the decoupling.  
According to the precise calculation, it is achieved when the universe 
expands about $e^{70}$ times during inflation \cite{LL,Riotto}.

\subsection{The origin of the large scale structure}
The fact that the comoving Hubble radius decreases during inflation 
makes it possible to generate the nearly scale-invariant density perturbations 
on large scales.  Since the scales of 
perturbations are within the Hubble radius in the early stage of inflation, 
causal physics works to generate small quantum fluctuations.  
After a scale is pushed outside the Hubble radius ({\it i.e.}, the first 
horizon crossing) during inflation, 
the perturbations can be described by classical ones.  
When the inflationary period ends, the evolution of the universe is followed by the 
standard big-bang cosmology, and the comoving Hubble radius begins to 
increase.  Then the scales of perturbations cross inside the Hubble radius 
again (the second horizon crossing), after which causality works.  The small 
perturbations imprinted during inflation appear as large-scale 
perturbations after the second horizon crossing.  In this way the 
inflationary paradigm naturally provides a causal mechanism in generating 
the seeds of density perturbations observed in the CMB anisotropies.

\subsection{Monopole problem}
During the inflationary phase ($\rho+3p<0$), the energy density of the 
universe decreases very slowly. 
For example, when the universe evolves as $a \propto t^p$ with $p>1$,
we have $H \propto t^{-1} \propto a^{-1/p}$ and $\rho \propto a^{-2/p}$.  
Meanwhile the energy density of massive particles decreases much faster 
($\sim a^{-3}$), these particles are red-shifted away during inflation, 
thereby solving the monopole problem.

Of course, we have to worry for the case where these unwanted particles are
produced {\it after} inflation.  In the process of reheating followed by 
inflation, the energy of the universe can be transferred to radiation or 
other light particles.  At this stage unwanted particles must not be 
overproduced in order not to violate the success of the standard cosmology 
such as nucleosynthesis.  Generally if the reheating temperature is 
sufficiently low, the thermal production of unwanted relics such as 
gravitinos can be avoided.

\section{Inflationary dynamics}

The scalar fields are the important ingredients in particle physics 
theories.  Consider a homogeneous scalar field, $\phi$,  
called {\it inflaton}, whose potential energy leads to the 
exponential expansion of the universe.  
The energy density and the pressure density of the inflaton can be described,
respectively, as 
\beqa 
\rho=\frac12\dot{\phi}^2+V(\phi)\,, 
~~~p=\frac12\dot{\phi}^2-V(\phi)\,,
\label{2_3_1}
\eeqa
where $V(\phi)$ is the potential of the inflaton.  Substituting 
Eq.~(\ref{2_3_1}) for Eqs.~(\ref{2_1_4}) and (\ref{2_1_5}), we get 
\beqa 
H^2=\frac{8\pi}{3m_{\rm pl}^2}
\left[\frac12 \dot{\phi}^2+V(\phi) \right]\,,
\label{2_3_2}
\eeqa
\beqa
\ddot{\phi}+3H\dot{\phi}+V'(\phi)=0\,,
\label{2_3_3}
\eeqa
where $\kappa^2 \equiv 8\pi G=8\pi/m_{\rm pl}^2$, 
and we neglected the curvature term $K^2/a^2$ in Eq.~(\ref{2_3_2}).

During inflation, the relation (\ref{2_2_2}) yields $\dot{\phi}^2<V(\phi)$,
which indicates that the potential energy of the inflaton dominates over the 
kinetic energy of it.  Therefore a flat potential of the inflaton is
required in order to lead to sufficient amount of inflation.  
Imposing the slow-roll conditions: $\frac12 \dot{\phi}^2 \ll V(\phi)$ 
and $\ddot{\phi} \ll 3H\dot{\phi}$, Eqs.~(\ref{2_3_2}) and (\ref{2_3_3}) 
are approximately given as 
\beqa 
H^2 \simeq \frac{8\pi}{3m_{\rm pl}^2}V(\phi),
\label{2_3_4}
\eeqa
\beqa
3H\dot{\phi}\simeq -V'(\phi).
\label{2_3_5}
\eeqa
Defining the so-called slow-roll parameters 
\beqa 
\epsilon \equiv 
\frac{m_{\rm pl}^2}{16\pi}\left(\frac{V'}{V}\right)^2\,,~~~ \eta \equiv 
\frac{m_{\rm pl}^2}{8\pi} \frac{V''}{V}\,,
\label{2_3_6}
\eeqa
we can easily verify that the above slow-roll approximations are valid when 
\beqa
\epsilon \ll 1, ~~~|\eta| \ll 1\,.
\label{2_3_7}
\eeqa
The inflationary phase ends when $\epsilon$ and $|\eta|$ grow of order unity.
A useful quantity to describe the amount of inflation is the number 
of e-foldings, defined by
\beqa
N \equiv \ln \frac{a_f}{a_i}=\int_{t_i}^{t_f} H dt\,,
\label{2_3_8}
\eeqa 
where the subscripts $i$ and $f$ denote the quantities at the beginning 
and the end of the inflation, respectively. 

In order to solve the flatness problem, $\Omega$ is required to be
$|\Omega_f-1|~\lsim~10^{-60}$ right after the end of inflation.
Meanwhile the ratio $|\Omega-1|$ between the initial and final 
phase of inflation is given by
\beqa
\frac{|\Omega_f-1|}{|\Omega_i-1|} \simeq \left(\frac{a_i}{a_f}
\right)^2=e^{-2N}\,,
\label{2_3}
\eeqa 
where we used the fact that $H$ is nearly constant during inflation.
Assuming that $|\Omega_i-1|$ is of order unity, the number of 
e-foldings is required to be $N~\gsim~70$ to solve the flatness problem.  We 
have similar number of e-foldings from the requirement to solve the horizon 
problem.

\section{Models of inflation} 

So far we did not mention the form of the inflaton potential, $V(\phi)$.  
Originating from the old inflationary scenario proposed by Guth \cite{Guth} 
and Sato \cite{Sato}, we now have varieties of inflationary models : $R^2$, 
new, chaotic, extended, power-law, hybrid, natural, supernatural, extranatural, 
eternal, D-term, F-term, brane, oscillating, trace-anomaly 
driven,..., etc.

These different kinds of models can be roughly classified in the following 
way \cite{Kolbclass}. The first class (type I) is the ``large field" model,
in which the initial value of the inflaton is large and it rolls down toward
the potential minimum at smaller $\phi$.
Chaotic inflation \cite{Linde83} is one of 
the representative models of this class.
The second class (type II) is the ``small field" model, in which 
the inflaton field is small initially and slowly evolves toward the potential 
minimum at larger $\phi$.  New inflation \cite{Linde82,AS82} and natural 
inflation \cite{Natural} are the examples of this type.  In the first model, 
the second derivative of the potential $V^{(2)}(\phi)$ usually takes 
positive values, whereas $V^{(2)}(\phi)$ can change its sign in the second 
model.  The third one (type III) is the hybrid (double) inflation model 
\cite{hybrid}, in which inflation typically ends by the phase transition 
triggered by the presence of the second scalar field (or by the second 
phase of inflation after the phase transition).

Let us briefly recap each type of inflationary models.

\subsubsection{Chaotic inflation--~an example of type I}

The chaotic inflation model is described by the quadratic or 
quartic inflaton potential,
\beqa
V(\phi)=\frac12 m^2\phi^2\,,~~~{\rm or}~~~
V(\phi)=\frac14 \lambda\phi^4\,.
\label{2_3_9}
\eeqa
The {\it ``chaotic"} means that initial conditions of 
inflaton are distributed chaotically.
According to this scenario, the region which undergoes the 
sufficient amount of inflation gives rise to our universe.

Let us investigate the evolution of the universe in the case of the 
quadratic potential.  
Then Eqs.~(\ref{2_3_4}) and (\ref{2_3_5}) read 
\beqa 
H^2 \simeq \frac{4\pi m^2\phi^2}{3m_{\rm pl}^2}\,,~~~~
3H\dot{\phi}+m^2\phi \simeq 0\,.
\label{2_3_10}
\eeqa
Combining these relations gives the following solutions:
\beqa
\phi \simeq \phi_i-\frac{m m_{\rm pl}}{2\sqrt{3\pi}}t\,,
\label{2_3_11}
\eeqa
\beqa
a \simeq a_i \exp \left[2\sqrt{\frac{\pi}{3}}\frac{m}{m_{\rm pl}} \left(\phi_i 
t-\frac{mm_{\rm pl}}{4\sqrt{3\pi}}t^2\right) \right]\,,
\label{2_3_12}
\eeqa
where $\phi_i$ is an integration constant corresponding to the initial 
values of the inflaton. 
We find from the relation (\ref{2_3_12}) that 
the universe expands exponentially during the initial stage of inflation.  
With the increase of the second term in the square bracket of 
Eq.~(\ref{2_3_12}), the expansion rate slows down.  Since the slow-roll 
parameters are expressed as 
\beqa 
\epsilon=\eta=\frac{m_{\rm pl}^2}{4\pi 
\phi^2}\,,
\label{2_3_13}
\eeqa
the inflationary period ends around 
$|\phi| \approx m_{\rm pl}/\sqrt{4\pi}$, after which 
the system enters a reheating stage.  The total amount of inflation is 
approximately expressed as 
\beqa 
N \simeq 2\pi \left(\frac{\phi_i}{m_{\rm pl}}\right)^2 -\frac12.
\label{2_3_14}
\eeqa
In order to lead to sufficient inflation $N~\gsim~70$,
we require the initial value to be $\phi_i~\gsim~3m_{\rm pl}$.  The inflaton 
mass, $m$, is constrained by the amplitude of density perturbations 
observed by the COBE satellite.  
In order to fit observations, $m$ is 
required to be \cite{LL} 
\beqa m \simeq 10^{-6} m_{\rm pl}\,.
\label{2_3_15}
\eeqa
In the quartic potential case, the self-coupling is constrained 
to be $\lambda \simeq 10^{-13}$ 
by the similar argument \cite{Linde,LL}.

\begin{figure}
\epsfxsize = 4.5in \epsffile{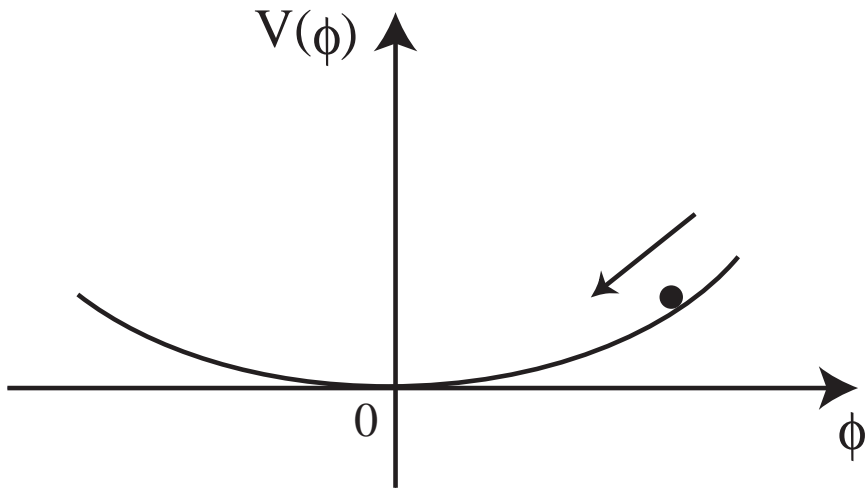} \caption{The schematic 
illustration of the potential of the chaotic inflation model.  This belongs 
to the class of the ``large field" model.}
\label{chaotic}
\end{figure}

\subsubsection{Natural inflation--~an example of type II}

Natural inflation model \cite{Natural} is characterized 
by Pseudo Nambu-Goldstone bosons (PNGBs) which appear
when an approximate global symmetry is 
spontaneously broken.
The PNGB potential is expressed as 
\beqa
V(\phi)=m^4 \left[1+\cos \left(\frac{\phi}{f}\right) \right]\,,
\label{2_3_16}
\eeqa
where two mass scales $m$ and $f$ characterize
the height and width of the potential, respectively. 
The typical mass scales are of order
$f \sim m_{\rm pl} \sim 10^{19}$ GeV and 
$m \sim m_{\rm GUT}\sim 10^{16}$ GeV,
which are expected by particle physics.

Let us consider the case where the inflaton is initially located in the 
region, $0 <\phi < \pi f$, and inflation occurs while the inflaton slowly 
evolves toward the potential minimum at $\phi=\pi f$.  The slow-roll 
parameters are 
\beqa \epsilon=\frac{m_{\rm pl}^2}{16\pi f^2}
\left[ \frac{\sin(\phi/f)}{1+\cos(\phi/f)} \right]^2,~~~~ 
\eta=-\frac{m_{\rm 
pl}^2}{8\pi f^2} \frac{\cos(\phi/f)}{1+\cos(\phi/f)}.
\label{2_3_17}
\eeqa
Note that $\epsilon$ and $\eta$ depend on $f$, but not on $m$.  
Inflation starts out from the regime where $\phi$ is close to zero.
The system enters a reheating stage when the inflaton 
begins to oscillate 
around $\phi =\pi f$.

One typical property in the type II model is that the second derivative of 
the inflaton potential can change its sign.  In the natural inflation, 
$V^{(2)}(\phi)$ is negative when inflaton evolves in the region of $0 <\phi 
< \pi f/2$.  This leads to the enhancement of inflaton fluctuations by 
spinodal (tachyonic) instability \cite{CH,TTspinodal,Felder}.  When the 
particle creation by spinodal instability is neglected, the number of 
$e$-foldings is expressed by \cite{Natural} 
\beqa 
N=\frac{16\pi f^2}{m_{\rm 
pl}^2} {\rm ln} \left[ \frac{\sin (\phi_f/2f)} {\sin (\phi_i/2f)} \right]\,,
\label{2_3_18}
\eeqa
where $\phi_i$ and $\phi_f$ are the initial and final values of the inflaton 
during inflation, respectively.  In order to obtain the sufficient amount of 
e-foldings satisfying $N~\gsim~70$, the initial value of the inflaton is 
required to be $\phi(t_i)~\lsim~0.1 m_{\rm pl}$ for the mass scale $f \sim 
m_{\rm pl}$.

\begin{figure}
\epsfxsize = 4.5in \epsffile{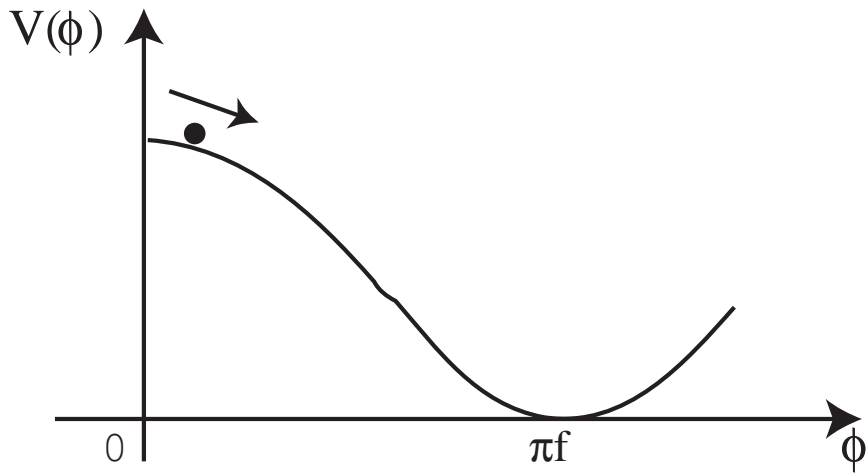} 
\caption{The schematic illustration of the potential
of the natural inflation model.  This belongs to the class of the ``small 
field" model.}
\label{natural}
\end{figure}

\subsubsection{Hybrid (Double) inflation--~an example of type III}

\begin{figure}
\epsfxsize = 4.5in \epsffile{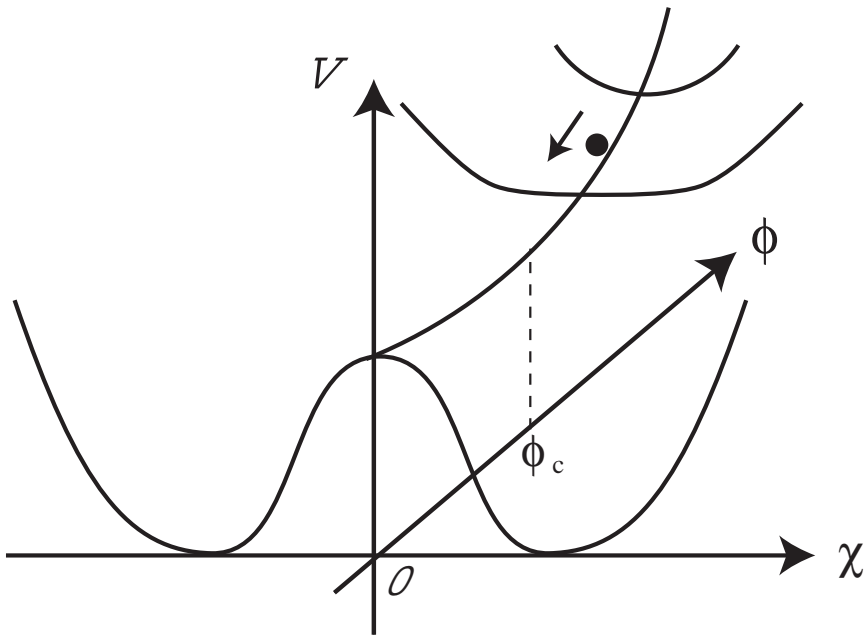} \caption{The schematic 
illustration of the potential of the hybrid (double) inflation model.  This 
model is characterized by multiple scalar fields.}
\label{hybrid}
\end{figure}

The inflationary model building in the presence of multiple scalar fields is 
the recent trend from the viewpoint of particle physics \cite{LR99}.  
Let us consider the Linde's hybrid inflation model \cite{hybrid}, 
described by 
\beqa 
V= \frac{\lambda}{4} \left(\chi^2-\frac{M^2} {\lambda}\right)^2 
+\frac12 g^2 \phi^2 \chi^2+ \frac12 m^2\phi^2\,.
\label{2_3_19}
\eeqa
When $\phi^2$ is large, the field tends to roll down toward the potential 
minimum at $\chi=0$.  In this case, the potential is effectively described 
by a single field, 
\beqa 
V \simeq \frac{M^4}{4\lambda} +\frac12 m^2\phi^2\,.
\label{2_3_20}
\eeqa
Inflation does not end for the potential (\ref{2_3_20}).
In the presence of the $\chi$ field, however, the mass 
of the field $\chi$ becomes negative for $\phi<\phi_c \equiv M/g$.  Then 
the field begins to roll down to one of the true minima at $\phi=0$ and 
$\chi=\pm M/\sqrt{\lambda}$ (see Fig.~\ref{hybrid}).  The original version 
of the hybrid inflation \cite{hybrid} corresponds to the case where the 
inflation soon comes to an end after the symmetry breaking ($\phi<\phi_c$) 
due to the rapid rolling of the field $\chi$.  
In this case the number of e-foldings acquired in double inflation 
can be approximately estimated by using the potential (\ref{2_3_20}): 
\begin{eqnarray}
N \simeq \frac{2\pi M^4} {\lambda m^2m_{\rm pl}^2}\,{\rm ln} 
\frac{\phi_i}{\phi_c} \,,
\label{N1}
\end{eqnarray}
where $\phi_i$ is the initial value of inflaton.

If the condition, $M^2 \gg 
\lambda M_p^2$, is satisfied, the mass of the field $\chi$ is ``light'' 
relative to the Hubble rate around $\phi=\phi_c$, thereby leading to the 
second stage of inflation for $\phi<\phi_c$ \cite{TPB}.  
This corresponds to the double inflationary scenario.

\section{Density perturbations from inflation}

In this section we study cosmological perturbations generated
from inflation.  For simplicity we shall consider the single-field model
of inflation.
The most general form of the line 
element that describes scalar metric perturbations is written as 
\cite{cosmo1,cosmo2} 
\begin{eqnarray}
ds^2=a^2(\tau)\{-(1+2A)d\tau^2+2B,_{i}dx^{i}d\tau 
+[(1-2\psi)\gamma_{ij}+2E,_{ij}]dx^idx^j\}\,,
\label{6_1_1}
\end{eqnarray}
where $\tau \equiv \int a^{-1} dt$ is the conformal time.  The scalar 
quantities $A$, $\psi$, $B$, and $E$ are the functions of space and time 
coordinates.  We denote the three-dimensional covariant derivative of a 
function $f$ with respect to some coordinate $i$ by $f_i$. 
In the linear perturbation theory of cosmological perturbations, 
there exist some redundant degress of freedom in the 
equations of motions which can be fixed without 
affecting the physics. 
It is important to choose a proper gauge condition so that it allows  for an 
simplified mathematical analysis and an easier physical interpretation 
(see {\em e.g.,} Ref.~\cite{hwang} for details).  Among several gauge 
conditions, the longitudinal gauge defined by the gauge condition $B=E=0$ 
is often used : 
\begin{eqnarray}
ds^2 &=&
a^2(\tau)[-(1+2\Phi) d\tau^2 +(1-2\Psi)\gamma_{ij}dx^idx^j] \nonumber \\
&=&
-(1+2\Phi)dt^2
+a^2(t)(1-2\Psi)\delta_{ij} dx^i dx^j\,.
\label{6_1_2}
\end{eqnarray}
The potentials $\Phi$ and $\Psi$ can be constructed as
gauge-invariant variables, whose forms  are invariant under
infinitesimal coordinate transformations (see Ref.~\cite{cosmo2} 
for details). When the spatial part of the energy-momentum tensor is diagonal, 
$\delta T^{i}_{j} \sim \delta^{i}_{j}$, $\Phi$ coincides with $\Psi$. 

Let us consider the linearized Einstein equation with $\Lambda=0$ : 
\begin{eqnarray}
\delta G^{\mu}_{\nu}=8\pi G \delta T^{\mu}_{\nu}.
\label{6_1_3}
\end{eqnarray}
The perturbed Einstein tensor can be evaluated in the longitudinal 
gauge as follows 
\begin{eqnarray}
\delta G^0_0 &=& 2a^{-2}\left[-3{\cal H}({\cal H} \Phi+\Psi') 
+\nabla^2 \Psi \right], \nonumber \\
\delta G^0_i &=& 2a^{-2}({\cal H} \Phi+\Psi') ,_{i}~, \nonumber \\
\delta G^i_j &=& -2a^{-2} \left\{\left[2{\cal H}'+{\cal H}^2)\Phi+ {\cal 
H}\Phi'+\Psi''+2{\cal H}\Psi'+\frac12 \nabla^2 D \right] \delta^i_j -\frac12 
\gamma^{ik} D,_{kj} \right\},
\label{6_1_4}
\end{eqnarray}
where ${\cal H} \equiv a'/a$ and $D \equiv \Phi-\Psi$, with a prime being a 
derivative with respect to $\tau$.  In the case where the universe is filled 
with a minimally coupled scalar field, $\phi$, the energy momentum tensor 
takes the following form 
\begin{eqnarray}
T^{\mu}_{\nu}=-(\nabla \phi)^2+\left[ \frac12
(\nabla \phi)^2+V (\phi) \right] \delta^{\mu}_{\nu}.
\label{6_1_5}
\end{eqnarray}
We decompose the field $\phi$ into homogenous and fluctuational parts as 
$\phi=\phi_0+\delta \phi$.  Then the energy momentum tensor is 
decomposed 
into background and perturbed parts as 
\begin{eqnarray}
T^{\mu}_{\nu}=^{(0)}T^{\mu}_{\nu}+\delta T^{\mu}_{\nu}\,,
\label{6_1_6}
\end{eqnarray}
where 
\begin{eqnarray}
^{(0)}T^0_0=\frac{1}{2a^2}\phi_0'^2+V(\phi_0)\,,~~~
^{(0)}T^0_i=0\,,~~~
^{(0)}T^i_j=\left[-\frac{1}{2a^2}\phi_0'^2+V(\phi_0)\right]
\delta^i_j\,,
\label{6_1_7}
\end{eqnarray}
and 
\begin{eqnarray}
\delta T^0_0 &=& a^{-2} \left[-\phi_0'^2\Phi+\phi_0' \delta\phi'+ 
V'(\phi)a^2\delta\phi \right]\,, \nonumber \\
\delta T^0_i &=& a^{-2} \phi_0'\delta\phi,_{i}\,, \nonumber \\
\delta T^i_j &=& a^{-2} \left[+\phi_0'^2\Phi-\phi_0' \delta\phi'+ 
V'(\phi)a^2\delta\phi \right]\delta^{i}_{j}\,.
\label{6_1_8}
\end{eqnarray}
Note that the relation $\delta T^i_j \sim \delta^i_j$ holds in this case.  This 
leads to the relation $D=\Phi-\Psi=0$ by considering the $ij$ part in
Eq.~(\ref{6_1_3}).  
Then the perturbed Einstein equation (\ref{6_1_3}) can be
written as 
\begin{eqnarray}
\nabla^2\Phi-3{\cal H}\Phi'-({\cal H}'+2{\cal H}^2)\Phi
=\frac{\kappa^2}{2}(\phi_0'\delta\phi+V'(\phi)a^2\delta\phi), 
\label{6_1_9}
\end{eqnarray}
\begin{eqnarray}
\Phi'+{\cal H}\Phi=\frac{\kappa^2}{2}\phi_0'\delta\phi, 
\label{6_1_10}
\end{eqnarray}
\begin{eqnarray}
\Phi''+3{\cal H}\Phi'+({\cal H}'+2{\cal H}^2)\Phi=
\frac{\kappa^2}{2}(\phi_0'\delta\phi-V'(\phi)a^2\delta\phi)\,,
\label{6_1_11}
\end{eqnarray}
where $\kappa^2 \equiv 8\pi G$. We used the background equations 
in deriving these equations.  Combining 
Eqs.~(\ref{6_1_9})-(\ref{6_1_11}) gives the equation for the 
gauge-invariant scalar field: 
\begin{eqnarray}
\delta\phi''+2{\cal H}\delta\phi'-\nabla^2\delta\phi+ 
V''(\phi)a^2\delta\phi =4\phi_0'\Phi'-2V'(\phi)a^2\Phi.
\label{6_1_12}
\end{eqnarray}
Note that the effect of gravitational perturbations appears in the 
rhs of Eq.~(\ref{6_1_12}). 

Eliminating the scalar field perturbation in 
Eqs.~(\ref{6_1_9})-(\ref{6_1_11}), we find the following equation for
gauge-invariant cosmological perturbations:
\begin{eqnarray}
\Phi''+2(a/\phi_0')'(a/\phi_0')^{-1}\Phi'-\nabla^2\Phi
+2\phi_0'({\cal H}/\phi')'\Phi=0\,.
\label{6_1_13}
\end{eqnarray}
Introducing new variables
\begin{eqnarray}
u=(a/\phi_0')\Phi\,,~~~~z={\cal H}/a\phi_0'\,,
\label{6_1_14}
\end{eqnarray}
Eq.~(\ref{6_1_13}) reduces to 
\begin{eqnarray}
u''-\nabla^2 u-(z''/z)u=0\,.
\label{6_1_15}
\end{eqnarray}
Making Fourier transformations, the mode function $u_k$ with 
a wave number $k$ satisfies\footnote{Note that the comoving wavelength 
$\lambda$ is related with $k$ by the relation $\lambda=2\pi/k$.} 
\begin{eqnarray}
u_k''+(k^2-z''/z)u_k=0\,.
\label{6_1_16}
\end{eqnarray}
In the long-wavelength limit, $k^2 \ll z''/z$, we find the analytic 
solution
\begin{eqnarray}
u=z \left(c_1+c_2 \int \frac{d\tau}{z^2}\right)\,,
\label{6_1_17}
\end{eqnarray}
where $c_1$ and $c_2$ are integration constants.  
Making use of Eqs.~(\ref{6_1_10}), (\ref{6_1_14}), and (\ref{6_1_17}), 
we have 
\begin{eqnarray}
\Phi=\frac{\dot{a}}{a^2}\left(c_1-2c_2 \int adt \right)+2c_2\,,
\label{6_1_18}
\end{eqnarray}
\begin{eqnarray}
\delta\phi=-\frac{\dot{\phi}_0}{a} \left(c_1-2c_2 \int adt \right)\,.
\label{6_1_19}
\end{eqnarray}
The solution (\ref{6_1_18}) indicates that there exists a 
gauge-invariant conserved quantity (comoving curvature perturbation)
in the long-wavelength limit: 
\begin{eqnarray}
{\cal R}=\Phi-\frac{H^2}{\dot{H}} \left(\Phi+\frac{\dot{\Phi}}{H} \right) 
=2c_2\,,
\label{6_1_20}
\end{eqnarray}
which was firstly introduced by Bardeen \cite{Bardeen}.

During inflation, the perturbations with a comoving wavenumber
$k$ are pushed outside the Hubble radius by the accelerated expansion
of the universe.  This makes the wavenumber toward the 
long wavelength regions: $k^2 \ll z''/z$, in which case we have the analytic 
solutions of Eqs.~(\ref{6_1_18}) and (\ref{6_1_19}).  Since the terms which 
include the $c_1$ coefficient rapidly decrease and the relation $|\dot{H}| 
\ll H^2$ holds during the inflationary phase, Eqs.~(\ref{6_1_18}) and 
(\ref{6_1_19}) yield 
\begin{eqnarray}
\Phi &\simeq& 2c_2 \left( \frac{1}{a} \int adt \right)^{\cdot{}} 
\nonumber \\
&=& 2c_2 \left[ H^{-1}-a^{-1} \int a(H^{-1})^{\cdot{}}dt 
\right]^{\cdot{}} \simeq -2c_2 \dot{H}/H^2\,,
\label{6_1_21}
\end{eqnarray}
\begin{eqnarray}
\delta\phi \simeq 2c_2 \dot{\phi}_0 a^{-1} \int adt \simeq 2c_2 
\dot{\phi}_0/H\,.
\label{6_1_22}
\end{eqnarray}
Combining Eqs.~(\ref{6_1_20}) and (\ref{6_1_22}), we have
\begin{eqnarray}
{\cal R}=\frac{H\delta\phi_k}{\dot{\phi}_0}\,,
\label{6_1_23}
\end{eqnarray}
where we introduced the subscript ``{\it k}" for the fluctuation, $\delta\phi$.
The $H\delta\phi_k/\dot{\phi}_0$ term should be evaluated when the 
wavelength of the perturbation crosses the Hubble radius: $k^2=z''/z$.  
Once the scales are pushed outside the Hubble radius during inflation, the 
evolution of perturbations can be described by Eqs.~(\ref{6_1_18}) and 
(\ref{6_1_19}).
Long after the inflation ends, the scale of the 
perturbations reenters the Hubble radius again, which generates the seeds 
for the galaxies and clusters.  Thus the inflationary paradigm naturally 
explains the origin of large-scale structure.

The spectrum of the comoving curvature perturbation is defined as 
\begin{eqnarray}
{\cal P}_{\cal R}=\frac{k^3}{2\pi^2}\langle|{\cal R}|^2 \rangle
=\frac{k^3}{2\pi^2}\frac{H^2}{\dot{\phi}_0^2}|\delta \phi_k|^2\,.
\label{spe}
\end{eqnarray}
{}From Eqs.~(\ref{6_1_21}) and (\ref{6_1_22}) we have 
$\Phi \simeq \epsilon H\delta\phi_k/\dot{\phi}_0$ with 
$\epsilon=-\dot{H}/H^2$ being the slow-roll parameter defined in 
eq.~(\ref{2_3_6}).  Taking note that $|4\phi_0'\Phi'| \ll 
|2V'(\phi)a^2\Phi|$ on super-Hubble scales and making use of the slow-roll 
relation $3H\dot{\phi} \simeq -V'(\phi)$, Eq.~(\ref{6_1_12}) can be 
approximately written as 
\begin{eqnarray}
\delta\phi_k''+2{\cal H}\delta\phi_k'+ 
\left[k^2+a^2V''(\phi)+6a^2\epsilon H^2 \right]\delta\phi_k=0\,.
\label{delp}
\end{eqnarray}
Redefining a field perturbation, $\delta \chi_k=a^{-1}\delta\phi_k$,
this equation reduces to
\begin{eqnarray}
\delta\chi_k''+\left[k^2-\frac{1}{\tau^2}\left(\nu^2-\frac14\right) \right] 
\delta\chi_k=0~~~{\rm with}~~~\nu^2=\frac94+9\epsilon-3\eta\,,
\label{delc}
\end{eqnarray}
where we used the fact that $a''/a=a^2(2-\epsilon)H^2 \simeq 
(2+3\epsilon)/\tau^2$.  When $\nu$ is a real number, the solution for this 
equation can be written in terms of the Hankel functions of the first and 
second kind: 
\begin{eqnarray}
\delta\chi_k=\sqrt{-\tau} \left[c_1(k) H_{\nu}^{(1)}(-k\tau)+ c_2(k) 
H_{\nu}^{(2)}(-k\tau) \right]\,.
\label{han}
\end{eqnarray}
Quantum fluctuations are generated in the ultraviolet regime ($k \gg aH)$,
which is required to take the plane-wave  form $e^{-ik\tau}/\sqrt{2k}$ for 
$-k\tau \gg 1$.  Matching Eq.~(\ref{han}) with this quantum fluctuation, we 
have $c_1(k)=\frac{\sqrt{\pi}}{2}e^{i(\nu+\frac12)\frac{\pi}{2}}$ and 
$c_2(k)=0$, thereby yielding 
\begin{eqnarray}
\delta\chi_k=\frac{\sqrt{\pi}}{2}e^{i(\nu+\frac12)\frac{\pi}{2}}
\sqrt{-\tau}H_{\nu}^{(1)}(-k\tau)\,.
\label{han2}
\end{eqnarray}
Since the Hankel function takes the form $H_\nu (-k\tau \ll 1) \simeq 
\sqrt{\frac{2}{\pi}}e^{-i\frac{\pi}{2}}2^{\nu-\frac32} 
(\Gamma(\nu)/\Gamma(3/2))(-k\tau)^{-\nu}$ on super-Hubble scales, the 
amplitude of the fluctuation $\delta\phi_k$ is estimated as 
\begin{eqnarray}
|\delta\phi_k| \simeq \frac{H}{\sqrt{2k^3}} 
\left(\frac{k}{aH}\right)^{\frac32-\nu}\,.
\label{del2}
\end{eqnarray}
Substituting this relation for eq.~(\ref{spe}), we finally get the spectrum of 
the curvature perturbations on super-Hubble scales:
\begin{eqnarray}
{\cal P}_{\cal R}=\frac{4\pi}{m_{\rm pl}^2 \epsilon} 
\left(\frac{H}{2\pi}\right)^2 \left(\frac{k}{aH}\right)^{3-2\nu} \equiv 
A_{\cal R}^2 \left(\frac{k}{aH}\right)^{n_{\cal R}-1}\,.
\label{power}
\end{eqnarray}
Here the spectral index $n_{\cal R}$ is given as 
\begin{eqnarray}
n_{\cal R}-1 \equiv \frac{d\,{\rm ln} {\cal P}_{\cal R}}{d\,{\rm ln} k} 
=-6\epsilon+2\eta\,.
\label{n}
\end{eqnarray}
When $n_{\cal R}=1$, the spectrum of the curvature perturbation is 
scale-invariant.  Since the slow-roll parameters $\epsilon$ and $\eta$
are much smaller than unity [see eq.~(\ref{2_3_7})], scalar
perturbations generated in standard inflation are close to scale-invariant.  
This property is consistent with the recent observation of WMAP 
\cite{Bennett:2003bz}.  When $n_{\cal R}<1$ and $n_{\cal R}>1$, we call the 
red-tilted spectrum and the blue-tilted spectrum, respectively.  For 
example the chaotic inflation model exhibits the red-tilted spectrum, while 
the hybrid inflation model shows the blue-tilted spectrum \cite{LL}.  This 
property is useful to constrain on inflationary models from observations.

The spectrum (\ref{power}) corresponds to the adiabatic perturbation
generated in the single-field inflationary scenario.  In the case of the 
multi-field inflation, this is modified due to the presence of 
isocurvature (entropy) perturbations \cite{Polarski:1994rz}.  In particular 
the curvature perturbation ${\cal R}$ is no longer a conserved quantity in 
the multi-field context \cite{Starobinsky:2001xq}.
Therefore we have to estimate the curvature perturbation at the end of
inflation instead of using its value at the horizon crossing.
See ref.~\cite{TPB} for the detailed numerical analysis of correlated 
adiabatic and isocurvature perturbations in concrete models of 
double inflation.

\section{Reheating after inflation}

After the end of inflation, the universe enters a reheating stage,
during which the potential energy of the inflaton is transferred to radiation 
and the universe is thermalized.  In the original version of the reheating 
scenario which is now called {\it old reheating}, the decay of the 
inflaton is described by the perturbation theory \cite{oldre}.  This process 
is not efficient for the success of the GUT scale baryogenesis scenario.  
In contrast, it was later found that the existence of the nonperturbative 
stage called {\it preheating} can lead to the explosive particle production 
in the early stage of reheating \cite{TB,KLS1}.  

\subsection{Old reheating}

The reheating stage turns on when the inflaton reaches the potential 
minimum and begins to oscillate.
Let us first study the background evolution during reheating for the  
polynomial potential 
\beqa 
V(\phi)=\frac{1}{2n}m^2m_{\rm pl}^2 
\left(\frac{\phi}{m_{\rm pl}}\right)^{2n}\,,
\label{3_1_4}
\eeqa
where $n$ is a positive integer.
Making use of the time-averaged relation 
\beqa 
\langle \dot{\phi}^2 \rangle_{T} = 2n 
\langle V(\phi) \rangle_{T}\,,
\label{taverage}
\eeqa
the time evolution of the scale factor and the 
Hubble parameter are 
\beqa 
a \propto t^{(n+1)/3n}\,,~~~H \propto \frac{n+1}{3n} 
\frac{1}{t}\,,
\label{3_1_5}
\eeqa
where we used eqs.~(\ref{2_3_2}) and (\ref{2_3_3}).
When $n=1$ and $n=2$, the universe 
evolves as matter-dominant ($a \propto t^{2/3}$) and radiation-dominant ($a 
\propto t^{1/2}$), respectively.

Let us consider the case of $n=1$ in order to understand 
the basic picture of reheating.
In this case, the evolution of the inflaton is described by the sinusoidal 
oscillation with a decreasing amplitude $\tilde{\Phi}(t)$: 
\beqa 
\phi=\tilde{\Phi}(t) \sin mt\,, ~~~
\tilde{\Phi}(t)=\frac{m_{\rm pl}} {\sqrt{3\pi}mt}\,.
\label{3_1_6}
\eeqa
Then the energy density of the inflaton decreases as 
\beqa 
\rho=\frac12 
\dot{\phi}^2+V(\phi) \approx \frac12 m^2
\tilde{\Phi}^2 \propto a^{-3}\,.
\label{3_1_7}
\eeqa
The simple estimations of Eqs.~(\ref{3_1_6}) and (\ref{3_1_7})
should be modified in the presence of the inflaton decay.
Without this decay, the universe will evolve toward
emptier and colder states because of the redshift of 
the energy density.
In the old reheating scenario \cite{oldre} where the single inflaton
decay is considered, the inflaton $\phi$ is coupled to 
a scalar field $\chi$ and a fermion field $\psi$ with the interacting
Lagrangian,
\beqa
{\cal L}_{\rm int}=-\sigma\phi\chi^2-h\phi\bar{\psi}\psi\,,
\label{3_1_8}
\eeqa
where $\sigma$ and $h$ are coupling constants.
Note that we assume that the bare masses of $\chi$ and $\psi$ 
are much smaller than the inflaton mass $m$ and neglect them. 
We include quantum correction effects to the equation of inflaton as 
\beqa 
\ddot{\phi}+3H\dot{\phi}+\left[m^2+\Pi(m)\right]\phi=0,
\label{3_1_9}
\eeqa
where $\Pi(m)$ is the polarization operator of inflaton.
Although the real part of $\Pi(m)$ is small relative to $m^2$,
$\Pi(m)$ acquires an imaginary part, 
\beqa 
{\rm Im}~\Pi(m)=-m\Gamma\,,
\label{3_1_10}
\eeqa
where $\Gamma=\Gamma(\phi \to \chi \chi)+
\Gamma(\phi \to \bar{\psi}\psi)$ is the total decay rate of inflaton.
The perturbative decay rates of $\phi \to \chi$ and $\phi \to \psi$
are expressed as \cite{oldre}
\beqa
\Gamma(\phi \to \chi \chi)=\frac{\sigma^2}{8\pi m}\,,~~~~
\Gamma(\phi \to \bar{\psi}\psi)=\frac{h^2m}{8\pi}\,.
\label{3_1_11}
\eeqa 
Note that these relations are valid in the case of $\Gamma \ll m$, 
which implies the relations: $\sigma^2 \ll m^2$ and $h^2 \ll 1$.  We cannot 
use Eq.~(\ref{3_1_11}) in the nonperturbative stage of reheating (preheating).

Combining Eqs.~(\ref{3_1_9}) and (\ref{3_1_10})
and setting $\phi=\tilde{\Phi}(t) \exp(imt)$, we find 
\beqa 
\tilde{\Phi}(t)=\frac{m_{\rm pl}}{\sqrt{3\pi}mt} \exp(-\Gamma t/2)\,.
\label{3_1_13}
\eeqa
The same solution can be obtained by adding the phenomenological 
decay term $\Gamma \dot{\phi}$ to the equation of inflaton:
\beqa
\ddot{\phi}+3H\dot{\phi}+\Gamma \dot{\phi}+m^2\phi=0\,.
\label{3_1_14}
\eeqa
The relation, $\Gamma<3H$, holds in the initial stage 
as long as coupling constants $\sigma$ and $h$
are small. Since the Hubble parameter decreases as $H \propto 1/t$,
the fraction of produced particles to the total energy density becomes
important when $3H$ drops less than $\Gamma$.
The energy density of the universe at this time can be estimated by 
setting $\Gamma^2=(3H)^2=24\pi \rho/m_{\rm pl}^2$, as
\beqa
\rho=\frac{\Gamma^2 m_{\rm pl}^2}{24\pi}\,.
\label{3_1_15}
\eeqa  
If we assume that this energy density is rapidly transferred to light 
particles and newly-produced particles acquire the radiation energy with 
temperature $T_R$ by instant thermalization, we obtain the relation 
\beqa 
\rho=\frac{g_* \pi^2 T_R^4}{30}=
\frac{\Gamma^2 m_{\rm pl}^2}{24\pi}\,,
\label{3_1_16}
\eeqa  
where $g_* (~\gsim~100)$ is the effective number of degrees of freedom
at $T=T_R$. Then the reheating temperature is estimated as
\beqa
T_R~\lsim~0.1 \sqrt{\Gamma m_{\rm pl}}\,.
\label{3_1_17}
\eeqa  
Taking into account the relation, 
$\Gamma \ll m~\lsim~10^{-6}m_{\rm pl}$, we find
\beqa
T_R \ll 10^{-4}m_{\rm pl}\,.
\label{3_1_18}
\eeqa  
This value is much smaller than the 
GUT scale temperature $T_{\rm GUT}=10^{-3}m_{\rm pl}$, 
which indicates that the GUT scale
baryogenesis does not work well in the old reheating scenario.

\subsection{Preheating}

In the early 1990s, it was realized that the inflaton decay may
have started in a much more explosive process, 
called {\it preheating} \cite{TB,KLS1} 
before the perturbative decay (see also refs.~\cite{Boy,KLS2,GKLS}).  
At this stage, the fluctuations of scalar particles can grow quasi-exponentially by 
parametric resonance.  The process of reheating consists of three stages: 
\vspace{0.2cm}

(i) Preheating stage where particles are nonperturbatively produced

(ii) Perturbative stage where the inflaton decay is described by 
the Born process

(iii) Thermalization of produced particles

\vspace{0.2cm}
Among variant models of inflation, chaotic inflation has an advantage 
that the dynamics of preheating is simply analyzed.
Kofman, Linde, and Starobinsky
investigated the structure of resonance in preheating 
both in massive and self-interacting inflationary 
models \cite{KLS1}.  
Hereafter we shall review the dynamics of 
preheating for the two-field model with a massive inflaton, 
\beqa 
V(\phi,\chi)=\frac12 m^2\phi^2+\frac12 g^2\phi^2\chi^2\,.
\label{3_2_1}
\eeqa  
We assume that the spacetime and the inflaton
$\phi$ give a classical background and the scalar field $\chi$
is treated as a quantum field on that background.
Expanding the scalar field $\chi$ as
\beqa
\chi=\frac{1}{(2\pi)^{3/2}} \int \left(a_k \chi_k(t)
 e^{-i {\bf k} \cdot {\bf x}}+a_k^{\dagger} \chi_k^{*}(t)
 e^{i {\bf k} \cdot {\bf x}} \right) d^3{\bf k}\,,
\label{3_2_2}
\eeqa  
and adopting the flat Friedmann-Robertson-Walker (FRW) metric, 
the each Fourier component $\chi_k(t)$ satisfies
the following equation of motion, 
\beqa 
\ddot{\chi}_k+3H\dot{\chi}_k+
\left(\frac{k^2}{a^2}+ g^2\phi^2 \right)\chi_k=0\,.
\label{3_2_3}
\eeqa  
Introducing a new scalar field $X_k \equiv a^{3/2}\chi_k$,
Eq.~(\ref{3_2_3}) yields
\beqa
\ddot{X}_k+\omega_k^2 X_k=0\,,
\label{3_2_4}
\eeqa  
where 
\beqa
\omega_k^2 \equiv \frac{k^2}{a^2} + g^2\phi^2
                     -\frac34 \left(\frac{2\ddot{a}}{a}+
                     \frac{\dot{a}^2}{a^2}\right)\,.
\label{3_2_5}
\eeqa 
We neglect the last term in Eq.~(\ref{3_2_5}) for analytic 
investigations since this term becomes gradually negligible 
during the reheating phase.

\begin{figure}
\epsfxsize = 4.5in 
\epsffile{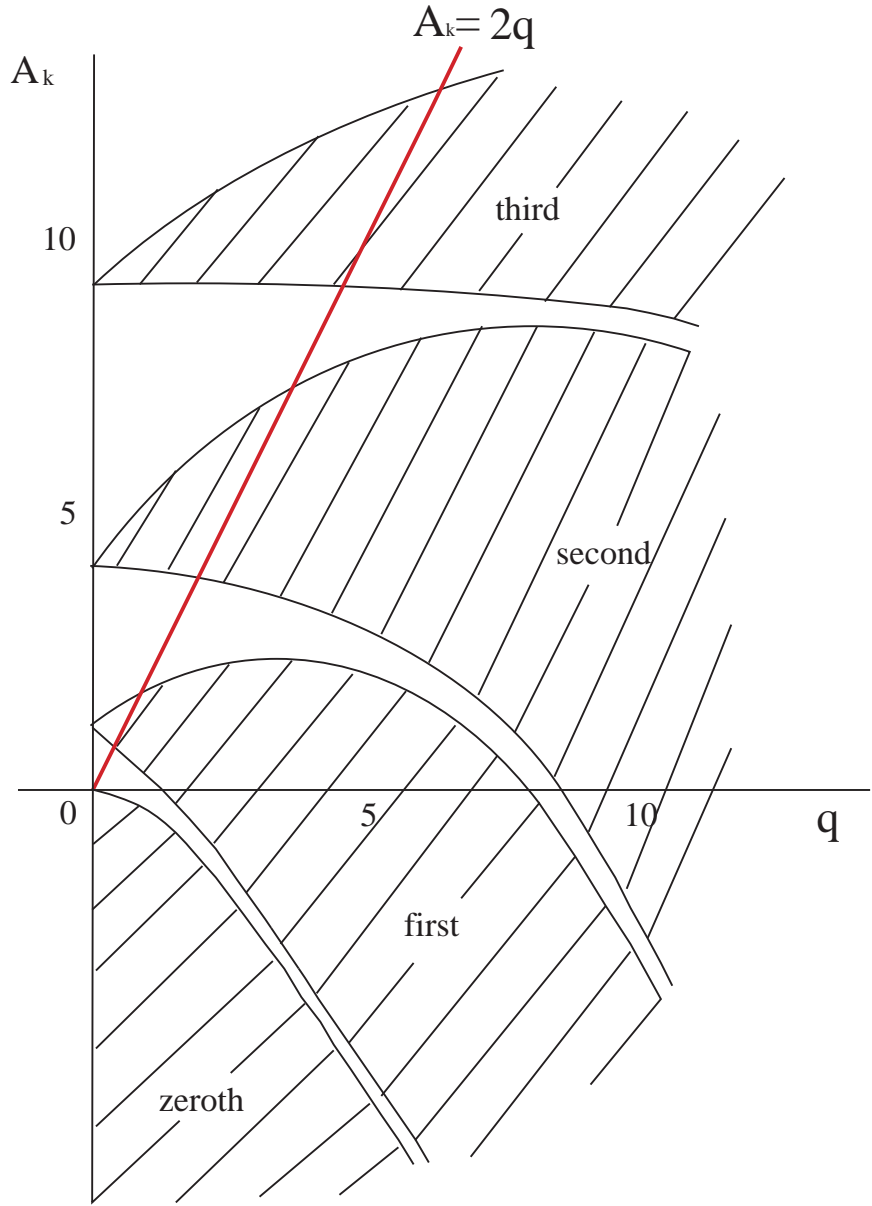} \caption{The schematic diagram of the 
Mathieu chart and the typical
paths for three types of resonance.  The lined regions
denote the  instability bands (zeroth, first, second,$\cdots$).
Note that the field perturbation evolves in the region $A_k \ge 2q$.}
\label{Mathieu}
\end{figure}

Then Eq.~(\ref{3_2_5}) is reduced to the well known 
Mathieu equation 
\begin{eqnarray}
\frac{d^2 X_k}{d z^2} + \left(A_k -2q \cos 2z \right) X_k=0\,,
 \label{3_2_6}
\end{eqnarray}
where $z=mt$ and
\begin{eqnarray}
A_k= 2q + \frac{k^2}{m^2a^2}\,,
 \label{3_2_7}
\end{eqnarray}
\begin{eqnarray}
q=\frac{g^2\tilde{\Phi}^2(t)}{4m^2}\,.
\label{3_2_8}
\end{eqnarray}
The strength of the resonance in Eq.~(\ref{3_2_6}) depends on the variables 
of $A_k$ and $q$, which is shown in a stability-instability chart of 
Fig.~\ref{Mathieu}.  In the unstable region (the lined region in 
Fig.~\ref{Mathieu}), $X_k$ grows exponentially as $X_k \propto {\rm exp} 
(\mu_k z)$ with the Floquet index $\mu_k$ and particles with momentum $k$ 
are produced.  For small $q~(\lsim~1)$, the width of the instability band 
is small and few $k$-modes grow by this resonance.  This is called the 
narrow resonance.  On the other hand, for the large $q (\gg 1)$, the 
resonance can occur for a broad range of the momentum $k$-space.  Since the 
growth rate of the $\chi$-particle becomes larger with the increase of the 
variable $q$, this resonance provides more efficient particle production 
than in the narrow one.  This is called the broad resonance \cite{KLS1}.  
Note that the initial amplitude of the inflaton and the coupling constant $g$ 
play important roles to determine whether the resonance is narrow or broad.  
Since the inflaton mass is constrained to be small as $m \sim 10^{-6}m_{\rm 
pl}$ by COBE normalizations, the large resonance parameter, $q \gg 1$, can 
be easily achieved for the coupling $g~\gsim~10^{-4}$ with the initial 
amplitude, $\tilde{\Phi}(t_i) \sim 0.2m_{\rm pl}$.  The allowed region on the 
Mathieu chart is determined by Eq.~(\ref{3_2_7}) as 
\begin{eqnarray}
A_k \geq 2q\,.
\label{3_2_9}
\end{eqnarray}
Hence the broadest resonance is given by the limit line $A_k =2q$.

As is found in Eq.~(\ref{3_2_7}), particles with low momenta are
mainly produced. Considering the nonadiabatic condition, 
$d\omega_k/dt \gg \omega_k^2$, for the particle production, the 
maximum comoving momentum is roughly estimated as \cite{KLS2}
\begin{eqnarray}
k~\lsim~\sqrt{\frac{g m \tilde{\Phi}}{2}}\,.
\label{3_2_9_1}
\end{eqnarray}
This indicates that large values of $g$ and $\tilde{\Phi}$ lead to the 
production of particles with high momenta.

When $q$ is sufficiently large initially ($q \gg 1$), 
the resonance of each mode occurs {\it stochastically} \cite{KLS2}.
In this case, the frequency $\omega_k$  decreases by cosmic expansion and 
$\omega_k$ drastically changes within each oscillation of the inflaton field, 
so the phases of $\chi$ field at successive moment of $\phi=0$ are not correlated each other. 
In the first stage of preheating, the $\chi$ fields cross large number of instability bands. 
The periods when they are in instability 
bands are so short that the resonance can not occur efficiently relative to  
the Minkowski spacetime case.
Nevertheless, the $\chi$ fluctuations can still grow quasi-exponentially.
As $q$ becomes smaller, cosmic expansion slows down, and the fields stay 
in each resonance band for a longer time.
When the variables 
decrease below the lower boundary of the first resonance band by the expansion 
of the universe, particle productions come to an end.

\begin{figure}
\epsfxsize = 4.5in 
\epsffile{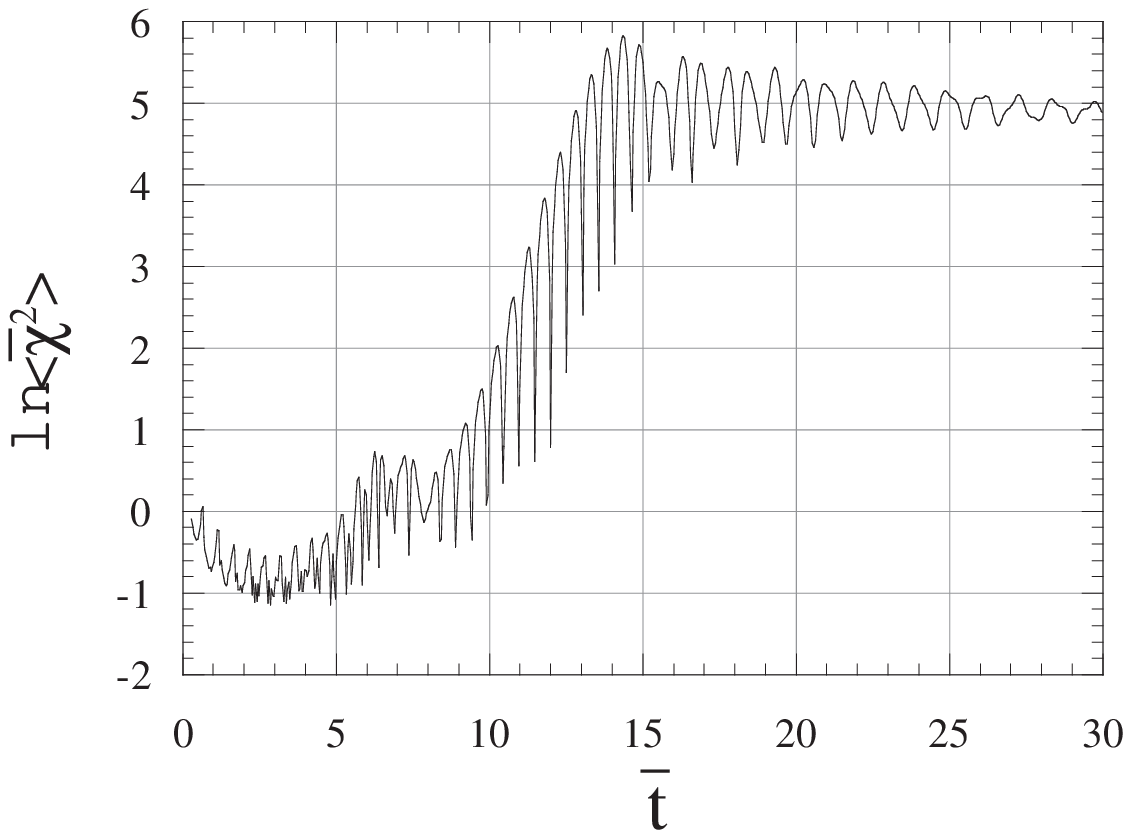} 
\caption{
The evolution of the variance 
$\langle\bar{\chi}^2\rangle=\langle\chi^2\rangle/m^2$ 
for $g=3.0 \times 10^{-4}$.}
\label{chievolution}
\end{figure}

We have to note  that there is another mechanism
which terminates the resonance.
Taking into account the backreaction effect of created $\chi$ particles,
the equation of inflaton is modified as
\beqa
\ddot{\phi}+3H\dot{\phi}+(m^2+g^2\langle\chi^2\rangle)\phi=0\,,
\label{3_2_11}
\eeqa
where the expectation value of $\chi^2$ is defined as
\beqa
\langle\chi^2\rangle \equiv \frac{1}{2\pi^2} \int k^2|\chi_k|^2~dk\,.
\label{3_2_12}
\eeqa 
When the initial value of $q$ is large as $q~\gsim~3000$,
which corresponds to $g~\gsim~3.0 \times 10^{-4}$,
the variance $\langle\chi^2\rangle$ grows of order $m^2/g^2$ 
and the backreaction onto the inflaton field cannot be ignored. 
This makes the oscillation of the inflaton  incoherent, which 
finally stops the resonance. 
The evolution of the variance $\langle\chi^2\rangle$ for $g=3.0 \times 
10^{-4}$ is shown in fig.~\ref{chievolution}.  

The growth of the field perturbations can stimulate the enhancement of 
metric perturbations on sub-Hubble and even on super-Hubble 
scales~\cite{metpre}.  
Whether this enhancement occurs or not is sensitive to 
the evolution of the large-scale $\chi$ fluctuation {\it during} inflation.  
In the model (\ref{3_2_1}), for example, strong amplification of the $\chi$ 
fluctuation requires a resonance parameter $q\gg 1$ at the beginning of 
preheating, in which case the large-scale $\chi_k$ modes are suppressed 
during inflation \cite{sup}.  Therefore it is difficult to amplify 
super-Hubble metric perturbations in this model.  However there exist 
several models of inflation where metric perturbations can be amplified 
during preheating \cite{TB2}.  This also provides an interesting possibility 
of overproducing primordial black holes in certain cases \cite{PBH}.

\section{Summary}

In this lecture note I reviewed the concept of inflation, generation of
density perturbations, and reheating after inflation.
I explained how inflationary cosmology solves a number of cosmological 
problems such as flatness, horizon, and monopole problems.  In addition 
inflation makes it possible to generate nearly scale-invariant density 
perturbations, which is consistent with observations.  Inflation is really 
an efficient mechanism to solve the cosmological problems associated
with standard big-bang cosmology.  In addition, elementary particles can be 
produced during the (p)reheating stage after inflation through the decay of 
the inflaton.

So far there exist several cosmological scenarios that might be the alternative 
to inflation such as the pre-big-bang \cite{Veneziano:1991ek} and 
ekpyrotic/cyclic model \cite{ekpyr}.  Nevertheless, unlike the standard 
inflation, it is generally difficult to solve all of the major cosmological 
problems at the same time in these models without a fine tuning of the model 
parameters.  It is fair to say that standard inflation with a slow-roll 
flat potential is the most promising scenario in the very early universe among 
the models proposed so far.

Nevertheless we still have unsolved problems even in the inflationary 
cosmology: What is the origin of the inflaton field?  What is the state of the 
universe before inflation?  The initial singularity can be avoided?

Fortunately we begin to get the high-precision observational data, which is 
expected to reveal the nature of inflation in details in near future. 
From the theoretical side, there are a lot of extensive works to try to 
construct viable models of inflation based on superstring and 
supergravity theories.  It is really encouraging that we are living 
in an exciting era where the early universe models can be strongly  
constrained from the upcoming observational data.

\section*{ACKNOWLEDGEMENTS}
I thank the organizers of the school, especially Burin Gumjudpai and 
Rachan Rangdee, for organizing nice school and conference.  



\begin{thebibliography}{99}  

\bibitem{Guth} 
A. H. Guth, Phys. Rev. D {\bf 23}, 347 (1981).

\bibitem{Sato} 
K. Sato, Mon. Not. R. Astron. Soc. {\bf 195}, 467 (1981);
Phys. Lett. {\bf 99B}, 66 (1981).

\bibitem{Star} 
A.~A.~Starobinsky,
Phys.\ Lett.\ B {\bf 91}, 99 (1980).

\bibitem{Linde82} 
A. Linde, Phys. Lett. {\bf 108B}, 389 (1982).

\bibitem{AS82}
A. Albrecht and P. Steinhardt, 
Phys. Rev. Lett. {\bf 48}, 1220 (1982).

\bibitem{Linde83} 
A. Linde, Phys. Lett. {\bf 129B}, 177 (1983).

\bibitem{Linde} 
A.  Linde, {\em Particle Physics and Inflationary Cosmology},
Harwood, Chur (1990).

\bibitem{LL} 
A.~R.~Liddle and D.~H. ~Lyth, {\em Cosmological inflation and
large-scale structure}, Cambridge University Press (2000).

\bibitem{LR99} 
D.  Lyth and A.  Riotto, Phys.  Rept. {\bf 314}, 1 (1999).

\bibitem{inper}
S.~W.~Hawking,
Phys.\ Lett.\ B {\bf 115}, 295 (1982); 
A.~A.~Starobinsky,
Phys.\ Lett.\ B {\bf 117}, 175 (1982); 
A.~H.~Guth and S.~Y.~Pi,
Phys.\ Rev.\ Lett.\  {\bf 49}, 1110 (1982); 
J.~M.~Bardeen, P.~J.~Steinhardt and M.~S.~Turner, 
Phys.\ Rev.\ D {\bf 28}, 679 (1983).

\bibitem{Bennett:2003bz}
C.~L.~Bennett {\it et al.},
arXiv:astro-ph/0302207;
H.~V.~Peiris {\it et al.},
arXiv:astro-ph/0302225.

\bibitem{Weinberg}
S.  Weinberg, {\em Gravitation and Cosmology},
J.~Wiley, New York (1972).

\bibitem{Gravitation}
C. W. Misner, K. S. Thorne, and J. A. Wheeler,  
{\em Gravitation}, W.~H.~Freeman, New York (1972).
 
\bibitem{Liddle}
A.~R.~Liddle, astro-ph/9612093 (1996).

\bibitem{Riotto}
A.~Riotto, hep-ph/0210162 (2002).
 
\bibitem{Kolbclass}
E. W. Kolb, hep-ph/9910311 (1999).

\bibitem{Natural}
K. Freese, J. A. Frieman, A. V. Orinto, Phys. Rev. Lett.
{\bf 65},  3233 (1990);
F. C. Adams, J. R. Bond, K. Freese, J. A. Frieman, and A. V. Orinto,
Phys. Rev. D {\bf 47}, 426 (1993).

\bibitem{hybrid}
A. D. Linde, Phys. Rev. D {\bf 49}, 748 (1994);
E. J. Copeland, A. R. Liddle, D.H. Lyth, E. D. Stewart,
and D. Wands, Phys. Rev. D {\bf 49}, 6410 (1994). 

\bibitem{CH}
D. Cormier and R. Holman, Phys. Rev. D {\bf 60}, 041301 (1999);
D. Cormier and R. Holman, Phys. Rev. D {\bf 62}, 023520 (2000).

\bibitem{TTspinodal}
S. Tsujikawa and T. Torii, Phys. Rev. D {\bf 62}, 043505 (2000).

\bibitem{Felder}
G. Felder et al., Phys. Rev. Lett. {\bf 87}, 011601 (2001);
Phys. Rev. D {\bf 64}, 123517 (2001). 

\bibitem{TPB}
S.~Tsujikawa, D.~Parkinson, B.~A.~Bassett,
astro-ph/ 0210322, {\it Physcial Review D to appear}.

\bibitem{cosmo1}
H. Kodama and M. Sasaki, Prog. Theor. Phys. Suppl.
No. 78, 1 (1984).

\bibitem{cosmo2}
V. Mukhanov, H. Feldman and R. Brandenberger,
Phys. Rep. {\bf 215}, 203 (1992).

\bibitem{hwang}
J.  Hwang, Astrophys.  J.  {\bf 375}, 443 (1991).

\bibitem{Bardeen}
J. Bardeen, Phys. Rev. {\bf D22}, 1882 (1980).

\bibitem{Polarski:1994rz}
D.~Polarski and A.~A.~Starobinsky,
Phys.\ Rev.\ D {\bf 50}, 6123 (1994);
M.~Sasaki and E.~D.~Stewart,
Prog.\ Theor.\ Phys.\  {\bf 95}, 71 (1996);
J.~c.~Hwang and H.~Noh,
Phys.\ Lett.\ B {\bf 495}, 277 (2000);
C.~Gordon, D.~Wands, B.~A.~Bassett and R.~Maartens,
Phys.\ Rev.\ D {\bf 63}, 023506 (2001).

\bibitem{Starobinsky:2001xq}
J.~Garcia-Bellido and D.~Wands,
Phys.\ Rev.\ D {\bf 53}, 5437 (1996);
A.~A.~Starobinsky, S.~Tsujikawa and J.~Yokoyama,
Nucl.\ Phys.\ B {\bf 610}, 383 (2001);
N.~Bartolo, S.~Matarrese and A.~Riotto,
Phys.\ Rev.\ D {\bf 64}, 123504 (2001).


\bibitem{oldre} 
A. Dolgov and A. Linde, Phys. Lett. {\bf 116B}, 329 
(1982); L. Abbott, E. Farhi, and M. Wise, Phys. Lett. {\bf 117B}, 29.

\bibitem{TB}
J. Traschen and R. H. Brandenberger, Phys. Rev. D {\bf 42}, 2491 (1990);
Y. Shatanov, J. Trashen, and R. H. Brandenberger, 
Phys. Rev. D {\bf 51}, 5438 (1995).

\bibitem{KLS1}
L. Kofman, A. Linde, and A. A. Starobinsky, Phys. Rev. Lett.
{\bf 73}, 3195 (1994).

\bibitem{Boy}
D. Boyanovsky, H. J. de Vega, R. Holman, D. S. Lee, and A. Singh,
Phys. Rev. D {\bf 51}, 4419 (1995);
M. Yoshimura, Prog. Theor. Phys. {\bf 94}, 873 (1995);
D. I. Kaiser, Phys. Rev. D {\bf 53}, 1776 (1995); 
S. Khlebnikov and I. I. Tkachev, Phys. Rev. Lett. {\bf 77}, 219 (1996);
J. Baacke, K. Heitmann, C. P\"atzold, Phys. Rev. D{\bf 55}, 2320 (1997).

\bibitem{KLS2}
L. Kofman, A. Linde, and A. A. Starobinsky, Phys. Rev. D {\bf 56},
3258 (1997).

\bibitem{GKLS}
P. B. Greene, L. Kofman, A. Linde, and A. A. Starobinsky,
Phys. Rev. D {\bf 56}, 6175 (1997).

\bibitem{metpre}
A. Taruya and Y. Nambu, 
Phys. Lett. {\bf B428}, 37 (1998);
F. Finelli and R.H. Brandenberger, 
Phys. Rev. Lett. {\bf 82}, 1362 (1999);
B.~A. Bassett, D.~I. Kaiser, and R. Maartens, 
Phys. Lett.  {\bf B455}, 84 (1999); 
M. Parry and R. Easther, Phys. Rev. D {\bf 62}, 
103503  (2000); B. A. Bassett and F. Viniegra, 
Phys. Rev. D {\bf 62}, 043507 (2000);
F. Finelli and R. H. Brandenberger, 
Phys. Rev. D {\bf 62}, 083502 (2000);
S.~Tsujikawa, B.~A. Bassett, and F. Viniegra, 
JHEP {\bf 08}, 019 (2000);
B.~A.~Bassett, M.~Peloso, L.~Sorbo and S.~Tsujikawa,
Nucl.\ Phys.\ B {\bf 622}, 393 (2002).

\bibitem{sup}
K. Jedamzik and G. Sigl, Phys. Rev. D {\bf 61}, 023519 (2000);
P. Ivanov, Phys, Rev. D {\bf 61}, 023505 (2000). 

\bibitem{TB2}
S.~Tsujikawa and B.~A.~Bassett,
Phys.\ Lett.\ B {\bf 536}, 9 (2002).

\bibitem{PBH}
A.~M. Green and K. A. Malik, 
Phys.  Rev.  D {\bf 64}, 021301 (2001);
B.~A. Bassett and S. Tsujikawa, 
Phys. Rev. D {\bf 63}, 123503 (2001);
F. Finelli and S. Khlebnikov, 
Phys.  Lett.  {\bf B 504}, 309 (2001).

\bibitem{Veneziano:1991ek}
G.~Veneziano,
Phys.\ Lett.\ B {\bf 265}, 287 (1991); 
M.~Gasperini and G.~Veneziano,
Astropart.\ Phys.\  {\bf 1} 317 (1993). 

\bibitem{ekpyr}
J.~Khoury, B.~A.~Ovrut, P.~J.~Steinhardt and N.~Turok,
Phys.\ Rev.\ D {\bf 64}, 123522 (2001);
P.~J.~Steinhardt and N.~Turok,
Phys.\ Rev.\ D {\bf 65}, 126003 (2002).

\end{thebibliography}
\end{document}